\documentclass[12pt,a4paper]{article}
\usepackage{epsfig}
\begin{document}
\title{Non-universal gauge bosons $Z^{\prime}$ and lepton flavor-violation tau decays }
\author{Chongxing Yue$^{(a,b)}$, Yanming Zhang$^{b}$, Lanjun Liu$^{b}$\\
{\small a: CCAST (World Laboratory) P.O. BOX 8730. B.J. 100080
        P.R. China}
\\ {\small b:College of Physics and Information Engineering,}\\
\small{Henan Normal University, Xinxiang  453002. P.R.China}
\thanks{E-mail:cxyue@public.xxptt.ha.cn} }
\date{\today}
\maketitle
\begin{abstract}
\hspace{5mm} There are many models beyond the standard model
predicting the existence of non-universal gauge bosons
$Z^{\prime}$ , which can give rise to  very rich phenomena. We
calculate the contributions of the non-universal gauge bosons
$Z^{\prime}$, predicted by topcolor-assisted technicolor (TC2)
models and flavor-universal TC2 models, to the lepton
flavor-violation tau decays $\tau\rightarrow l_{i}\gamma$ and
$\tau\rightarrow l_{i}l_{j}l_{k}$. We find that the branching
ratio $B_{r}(\tau\longrightarrow l_{i}l_{j}l_{k})$ is larger than
that of the process $\tau\longrightarrow l_{i}\gamma$ in all of
the parameter space. Over a sizable region of the parameter space,
we have $B_{r}(\tau \longrightarrow l_{i}l_{j}l_{k})\sim 10^{-8}$,
which may be detected in the future experiments.
\end {abstract}

\newpage

  Although the standard model (SM) has been successful in describing
 the physics of  electroweak interactions, it is quite possible that
the SM is only an effective theory valid below some high energy
scale. Extra gauge bosons $Z^{\prime}$ are the best motivated
extensions of the SM. If discovered they would represent
irrefutable proof of new physics, most likely that the SM gauge
groups must be extended\cite{y1}. If these extensions are
associated with flavor symmetry breaking, the gauge interactions
will not be flavor-universal\cite{y2}, which predict the existence
of non-universal gauge bosons $Z^{\prime}$. Furthermore,
 the possible anomalies  in the $Z$ pole $b\bar{b}$ asymmetries may
suggest a non-universal $Z^{\prime}$\cite{y3}.

After the mass diagonalization from the flavor eigenbasis into the
mass eigenbasis, the non-universal gauge interactions result in
the tree level flavor-changing neutral currents (FCNC's)
couplings. Thus, the non-universal gauge bosons $Z^{\prime}$ may
have significant contributions to some FCNC processes. For
example, the non-universal gauge bosons $Z^{\prime}$ predicted by
strong top dynamical models, such as topcolor-assisted technicolor
(TC2) models\cite{y4} and flavor-universal TC2 models\cite{y5},
can give significant contributions to some FCNC
processes\cite{y6,y7,y8}. If these effects are indeed detected at
LHC, LC or other experiments, it will be helpful to identify the
gauge bosons $Z^{\prime}$, and hence unravel underlying
theory\cite{y9}.

The high statistic results of the SuperKamiokande (SK) atmospheric
neutrino experiment \cite{y10} and the solar neutrino
experiment\cite{y11} have made one believe that neutrinos are
massive and oscillate in flavor and are of interest in the lepton
flavor violation (LFV) processes, such as $l_{i}\rightarrow
l_{j}\gamma$ and $l_{i}\rightarrow l_{j}l_{k}l_{l}$. These LFV
processes are practically suppressed to zero in the SM, due to the
unitarity of the leptonic analog of CKM mixing matrix and the near
masslessness of the three neutrinos. Thus, the observation of any
rate for one of these processes would be a signal of new physics.
The improvement of their experimental measurements forces one to
make more elaborate theoretical calculation in the framework of
some specific models beyond the SM and see whether the LFV effects
can be tested in the future experiments. For instance, the LFV tau
decays, such as $\tau\rightarrow \mu\gamma$, $\tau\rightarrow
e\gamma$, $\tau\rightarrow 3\mu$ and $\tau\rightarrow 3e$, have
been studied in a  model independent way in Ref.[12], in the SM
with extended right-handed and left-handed neutrino sectors
\cite{y13}, in supersymmetric  models\cite{y14}, in the two Higgs
doublets model type III \cite{y15} and in Zee model\cite{y16}.

In this letter, we focus our attention on the LFV decays of tau
lepton.  Now the bounds on the LFV tau decay modes are being
improved by the Belle experiment at KEK. The new bounds are
\cite{y17}:
$$B_{r}(\tau\longrightarrow \mu\gamma)< 5\times10^{-7},$$
$$B_{r}(\tau\longrightarrow    3\mu )< 8.7\times10^{-7},$$
$$B_{r}(\tau\longrightarrow   2\mu e)< 7.7\times10^{-7},$$
$$B_{r}(\tau\longrightarrow    2e\mu)< 3.4\times10^{-7},$$
$$B_{r}(\tau\longrightarrow   3 e)< 7.8\times10^{-7}.$$
We will calculate the contributions of the non-universal gauge
bosons $Z^{\prime}$, predicted by TC2 models and flavor-universal
TC2 models, to the branching ratios $B_{r}(\tau\rightarrow
l_{i}\gamma)$ and $B_{r}(\tau\rightarrow l_{i}l_{j}l_{k})$. It
will be shown that the branching ratios of the processes
$\tau\rightarrow l_{i}l_{j}l_{k}$, which occur at the tree level
are much larger than those of the processes $\tau\rightarrow
l_{i}\gamma$, which arise from$\gamma$-penguins with the photon on
the mass shell. With reasonable values of the parameters, the
values of $B_{r}(\tau\rightarrow l_{i}l_{j}l_{k})$ can reach
$4.1\times10^{-8}$, which may be tested by the near future
experiments. For the branching ratio $B_{r}(\tau\rightarrow
l_{i}\gamma)$, it is possible to reach $2.6\times10^{-9}$.

To completely avoid the problems, such as triviality and
unnaturalness arising from the elementary Higgs in the SM, various
kinds of dynamical models are proposed and among which TC2 models
are very interesting because they explain the large top quark mass
and provide possible dynamics of electroweak symmetry breaking
(EWSB)\cite{y6}. A common feature of these models is that the SM
gauge groups are extended at energy well above the weak scale.
Breaking of the extended gauge groups to their diagonal subgroups
produces non-universal massive gauge bosons. For instance, TC2
models and flavor-universal TC2 models all predict the existence
of the non-universal gauge bosons $Z^{\prime}$. These new
particles treat the third generation fermions differently from
those in the first and second generations. Thus, they can lead to
FC couplings.

The flavor-diagonal couplings of $Z'$ to leptons can be written as
\cite{y4,y6}:
\begin{eqnarray}
{\cal L}^{FD}_{Z'}=&-&\frac{1}{2}g_{1}cot\theta'
Z'_{\mu}(\bar{\tau}_{L}\gamma^{\mu}\tau_{L}+2\bar{\tau}_{R}\gamma^{\mu}\tau_{R})\nonumber\\
&-&\frac{1}{2}g_{1}tan\theta'Z'_{\mu}(\bar{\mu}_{L}\gamma^{\mu}\mu_{L}
+2\bar{\mu}_{R}\gamma^{\mu}\mu_{R}+\bar{e}_{L}\gamma^{\mu}e_{L}
+2\bar{e}_{R}\gamma^{\mu}e_{R}),
\end{eqnarray}
where $g_{1}$ is the ordinary hypercharge gauge coupling constant,
$\theta'$ is the mixing angle with
$tan\theta'=\frac{g_{1}}{\sqrt{4\pi k_{1}}}$. To obtain the top
quark condensation and not form a $b\overline{b}$ condensation,
there must be $tan\theta'<<1$\cite{y4,y5}. The flavor changing
couplings of $Z'$ to leptons can be written as:
\begin{eqnarray}
{\cal
L}^{FC}_{Z'}=&-&\frac{1}{2}g_{1}Z'_{\mu}[k_{\tau\mu}(\bar{\tau}_{L}\gamma^{\mu}\mu_{L}
+2\bar{\tau}_{R}\gamma^{\mu}\mu_{R})+k_{\tau
e}(\bar{\tau}_{L}\gamma^{\mu}e_{L}
+2\bar{\tau}_{R}\gamma^{\mu}e_{R})\nonumber\\
&+&k_{\mu
e}tan^{2}\theta(\bar{\mu}_{L}\gamma^{\mu}e_{L}+2\bar{\mu}_{R}\gamma^{\mu}e_{R})],
\end{eqnarray}
where $k_{ij}$ are the flavor mixing factors. In the following
estimation, we will take $k_{\tau\mu}=k_{\tau e}=k_{\mu
e}=k=\lambda$\cite{y6}, where $\lambda=0.22$ is the Wolfenstein
parameter\cite{y18}.

From Eq.(1) and Eq.(2), one can see that the LFV tau decays can be
generated via exchange of $Z'$ in the TC2 models and
flavor-universal TC2 models. We first consider the LFV decay
processes $\tau\rightarrow\mu\gamma$ and $\tau\rightarrow
e\gamma$. These processes are generated by the on-shell photon
penguin diagrams Fig.1($a$). The internal fermion line may be
$\tau$, $\mu$ or $e$. However, the two internal $\tau$ propagators
provide a term proportional to $m^{2}_{\tau}$ in the numerator,
which is not cancelled by the $m^{2}_{\tau}$ in the denominator
since the heavy $Z'$ boson mass dominate the denominator. Thus, we
ignore the contributions of the LFV couplings $Z'\mu e$,
$Z'\mu\mu$ and $Z'ee$ and only consider those of the LEV couplings
$Z'\tau\tau$, $Z'\tau\mu$ and $Z'\tau e$. After straightforward
calculation, we can derive the widths of
$\tau\rightarrow\mu\gamma$ and $\tau\rightarrow e\gamma$, which
are\cite{y19}:
\begin{equation}
\Gamma(\tau\rightarrow\mu\gamma)=\Gamma(\tau\rightarrow
e\gamma)=\frac{\alpha^{2}k_{1}}{1152\pi^{2}C^{2}_{W}}\frac{m^{5}_{\tau}}{M^{4}_{Z}}\cdot
k^{2},
\end{equation}
where $C^{2}_{W}=cos^{2}\theta_{W}$, $\theta_{W}$ is the Weinberg
angle and $M_{Z}$ is the mass of the non-universal gauge bosons
$Z'$. The non-universal gauge bosons $Z'$ can contribute to the
processes $\tau\rightarrow l_{i}l_{j}l_{k}$ via the Feynman
diagrams as depicted in Fig1.($b$) and Fig1.($c$). The
contributions of the off-shell photon penguin and $Z$ penguin to
these processes are much smaller than those of the $Z'$ exchange
at tree level. Thus we ignore the contributions of Fig.2(b) to
$\tau\rightarrow l_{i}l_{j}l_{k}$ in our calculation. The partial
widths can be written  as\cite{y20}:
\begin{eqnarray}
\Gamma(\tau\rightarrow 3\mu)&=&\Gamma(\tau\rightarrow
3e)=\Gamma(\tau\longrightarrow
ee\mu)\nonumber\\
&=&\Gamma(\tau\rightarrow\mu\mu e)=\frac{25\alpha^{3}}{384\pi
k_{1}C_{W}^{6}}\frac{m_{\tau}^{5}}{M_{Z}^{4}}\cdot k^{2}
\end{eqnarray}

The width of $\tau\longrightarrow\mu\gamma$ is equal to that of
$\tau\rightarrow e\gamma$ and the widths of the process
$\tau\rightarrow l_{i}l_{j}l_{k}$ are equal each other. This is
because the non-universal gauge bosons $Z'$ only treat the
fermions in the third generation differently from those in the
first and second generation and treat the fermions in the first
generation same as those in the second generation. The
$e\bar{\nu}_{e}\nu_{\tau}$ is one of the dominant decay modes of
the lepton $\tau$. The branching ratio $B_{r}(\tau\rightarrow
e\bar{\nu}_{e}\nu_{\tau})$ has been precisely measured, i.e.
$B_{r}(\tau\rightarrow e\bar{\nu}_{e}\nu_{\tau})=(17.83\pm
0.06)\%$ \cite{y21}. Thus, we can use $B_{r}(\tau\rightarrow
e\bar{\nu}_{e}\nu_{\tau})$ to represent the branching ratios of
the LFV tau decays:
\begin{equation}
B_{r}(\tau\longrightarrow l_{i}\gamma)=B_{r}(\tau\rightarrow
e\bar{\nu}_{e}\nu_{\tau})\frac{\Gamma(\tau\rightarrow
l_{i}\gamma)}{\Gamma(\tau\rightarrow e\bar{\nu}_{e}\nu_{\tau})},
\end{equation}
\begin{equation}
B_{r}(\tau\longrightarrow l_{i}l_{j}l_{k})=B_{r}(\tau\rightarrow
e\bar{\nu}_{e}\nu_{\tau})\frac{\Gamma(\tau\rightarrow
l_{i}l_{j}l_{k})}{\Gamma(\tau\rightarrow
e\bar{\nu}_{e}\nu_{\tau})},
\end{equation}
with
\begin{equation}
\Gamma(\tau\longrightarrow
e\bar{\nu}_{e}\nu_{\tau})=\frac{m_{\tau}^{5}G_{F}^{2}}{192\pi^{3}}.
\end{equation}
Here the Fermi coupling constant $G_{F}=1.16639\times
10^{-5}GeV^{-2}$\cite{y21}.

It has been shown that vacuum tilting and the constraints from
$Z$-pole physics and $U(1)$ triviality require $k_{1}\leq
1$\cite{y5}. The limits on the mass of $Z^{\prime}$ can be
obtained via studying its effects on various experimental
observables\cite{y6}. For example, Ref.[2] has shown that to fit
the electroweak measurement data, the $Z^{\prime}$ mass must be
larger than $1TeV$. As numerical estimation, we take the
$Z^{\prime}$ mass $M_{Z}$ and $k_{1}$ as free parameters.

In Fig.2 and Fig.3 we plot the branching ratios
$B_{r}(\tau\rightarrow l_{i}\gamma)$ and $B_{r}(\tau\rightarrow
l_{i}l_{j}l_{k})$ as functions of $M_{Z}$ for three values of the
parameter $k_{1}$: $k_{1}=0.2$, $0.5$ and $1$, respectively. From
Fig.2 and Fig.3 we can see that the branching ratio
$B_{r}(\tau\rightarrow l_{i}l_{j}l_{k})$ is larger than that of
the process $\tau\rightarrow l_{i}\gamma$ in all of the parameter
space. The values of $B_{r}(\tau\rightarrow l_{i}\gamma)$ increase
with $k_{1}$ increasing, while the values of
$B_{r}(\tau\rightarrow l_{i}l_{j}l_{k})$ decrease with $k_{1}$
increasing. For $k_{1}=1$, $M_{Z}=1TeV$, the value of
$B_{r}(\tau\rightarrow l_{i}l_{j}l_{k})$ can reach maximum value
i.e. $B_{r}(\tau\rightarrow l_{i}l_{j}l_{k})=4.1\times 10^{-8}$.
For $B_{r}(\tau\longrightarrow l_{i}\gamma)$, the  maximum value
is $2.6\times 10^{-9}$.

 The existence of non-universal gauge bosons
$Z^{\prime}$ is a key feature of all TC2 models. The new gauge
bosons can lead to the flavor-changing couplings, and hence may
have significant contributions to some FCNC processes. In this
letter, we calculated the contributions of non-universal gauge
bosons $Z^{\prime}$ predicted by TC2 models and flavor-universal
TC2 models to the LFV $\tau$ decays. We found that, in most of
parameter space, the branching ratio $B_{r}(\tau\longrightarrow
l_{i}\gamma)$ is smaller than $2.6\times 10^{-9}$, which is very
difficult to be tested. However, the branching ratio
$B_{r}(\tau\longrightarrow l_{i}l_{j}l_{k})$ can be larger over a
wide range of parameter space, $ B_{r}(\tau\longrightarrow
l_{i}l_{j}l_{k}) \sim 10^{-8}$. Thus, the effects of non-universal
gauge bosons $Z^{\prime}$ on the tau decays $\tau\longrightarrow
l_{i}l_{j}l_{k}$ may be detected in the near future experiments.

\vspace{1.5cm} \noindent{\bf Acknowledgments}

Chongxing Yue thanks the Abdus Salam International Centre  for
Theoretical Physics (ICTP) for partial support. We thank the
referee for carefully reading the manuscript. This work was
supported in part by the National Natural Science Foundation of
China (I9905004) and Foundation of Henan Educational Committee.

\newpage
\begin{center}
{\bf Figure captions}
\end{center}
\begin{description}
\item[Fig.1:] Feynman diagrams for lepton flavor-violation $\tau$
decays $\tau\rightarrow l_{i}\gamma$ and $\tau\rightarrow
l_{i}l_{j}l_{k}$ induced by non-universal gauge bosons $Z'$
exchange.
\item[Fig.2:]Branching ratios for $\tau\longrightarrow
l_{i}\gamma$ as functions of the gauge bosons $Z'$ mass $M_{Z}$
for $k_{1}=0.2$ (solid line), 0.5(dashed line) and 1 (dotted
line).
\item[Fig.3:] Branching ratios of $\tau\rightarrow l_{i}l_{j}l_{k}$
as functions of the gauge bosons $Z'$ mass $M_{Z}$ for $k_{1}=0.2$
(solid line), 0.5(dashed line) and 1 (dotted line).

\end{description}

\newpage
\begin{figure}[hb]
\begin{center}
\begin{picture}(250,400)(0,0)
\put(-50,0){\epsfxsize120mm\epsfbox{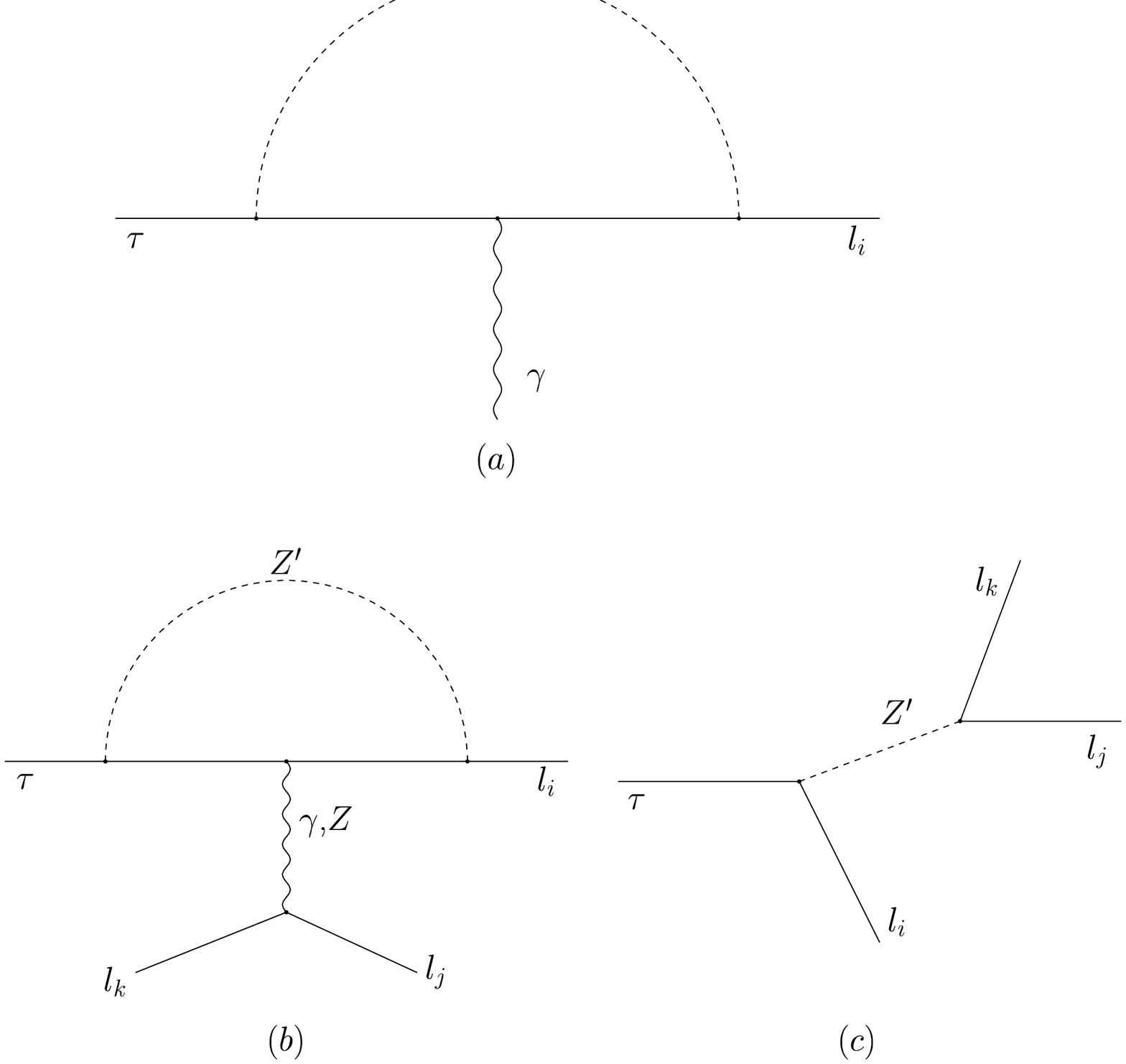}}
 \put(120,100){Fig.1}
\end{picture}
\end{center}
\end{figure}

\begin{figure}[hb]
\begin{center}
\begin{picture}(250,200)(0,0)
\put(-50,60){\epsfxsize 120mm \epsfbox{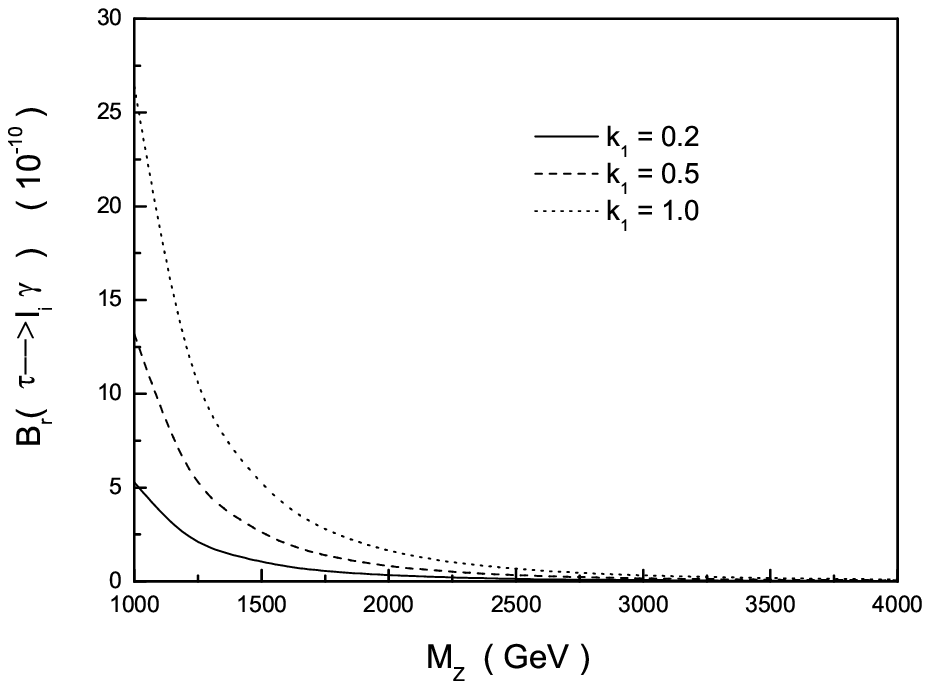}}
\put(120,40){Fig.2}
\end{picture}
\end{center}
\end{figure}

\begin{figure}[hb]
\begin{center}
\begin{picture}(250,200)(0,0)
\put(-50,50){\epsfxsize 120mm \epsfbox{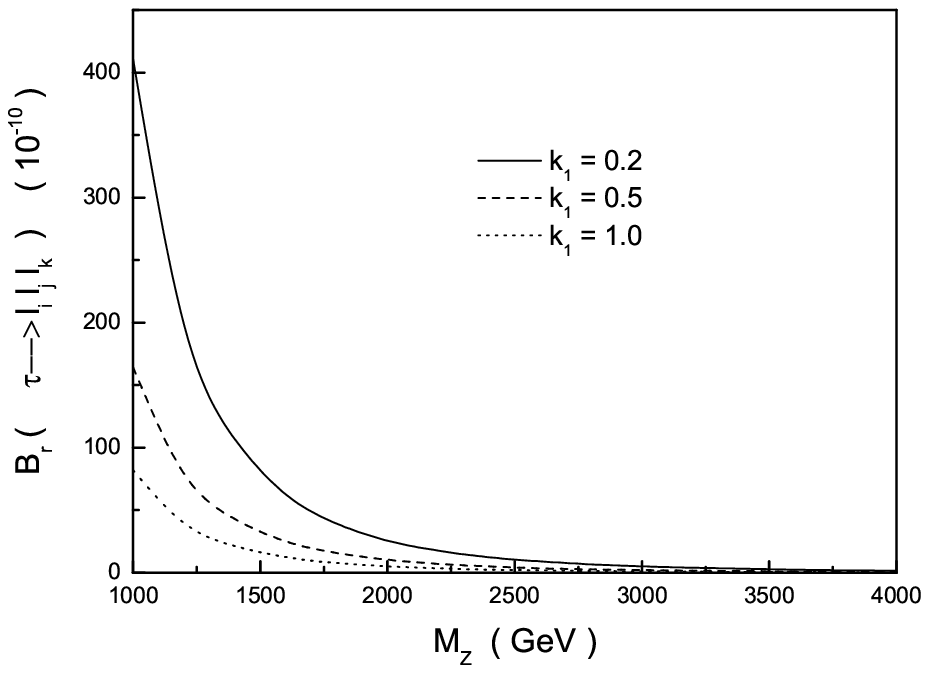}}
\put(120,30){Fig.3}
\end{picture}
\end{center}
\end{figure}

\end{document}